\documentclass[12pt,reqno]{amsart}
\usepackage{graphicx}  % Add graphics capabilities
\usepackage{amsmath}  % Better maths support & more symbols
\usepackage{amssymb,amsthm}
\usepackage{color}
\usepackage{tikz}
\usepackage{pgfplots}
\usetikzlibrary{decorations.pathmorphing}

\newcommand{\R}{{\mathbb R}}\newcommand{\N}{{\mathbb N}}

\newcommand{\Z}{{\mathbb Z}}\newcommand{\C}{{\mathbb C}}

\let\epsilon\varepsilon
\let\theta\vartheta
\let\hat\widehat

\newtheorem{theorem}{Theorem}[section]\newtheorem{lemma}[theorem]{Lemma}

\newtheorem{remark}[theorem]{Remark}

\title[KP-II approximation for a scalar FPU system]{KP-II approximation for a scalar FPU system \\	on a 2D square lattice}
	
\author[D. E.  Pelinovsky]{Dmitry E.Pelinovsky}
\address{ Department of Mathematics and Statistics, McMaster University, Hamilton,
	Ontario, Canada, L8S 4K1}
\email{dmpeli@math.mcmaster.ca}

\author[G. Schneider]{ Guido Schneider}
\address{Institut f\"ur Analysis, Dynamik und Modellierung, Universit\"at Stuttgart, 70569 Stuttgart, Germany}
\email{guido.schneider@mathematik.uni-stuttgart.de}

\begin{document}
	
\maketitle
	
	\begin{abstract}
		We consider a scalar Fermi--Pasta--Ulam (FPU) system on a square 2D lattice. The Kadomtsev--Petviashvili (KP-II) equation can be derived 
		by means of multiple scale expansions to describe unidirectional long waves of small amplitude with slowly varying  transverse modulations. We show that the KP-II approximation makes correct predictions about the dynamics of the original scalar FPU system. An existing approximation result is extended to an arbitrary direction of wave propagation. The main novelty of this work is the use of Fourier transform in the analysis of the FPU system in strain variables.
	\end{abstract}

\section{Introduction}

\begin{figure}[htpb!]
	\centering
	\begin{tikzpicture}[
	wall/.style = {gray,fill=gray},
	mass/.style = {draw,circle,ball color=red},
	spring/.style = {decorate,decoration={zigzag, pre length=.3cm,post length=.3cm,segment length=#1}},
	]
	\coordinate (l1) at (0,2);
	\node[mass,label={[xshift=0.9cm, yshift=0.05cm]$_{m-1,n-1}$}] (m11) at (2,2) {};
	\node[mass,label={[xshift=0.9cm, yshift=0.05cm]$_{m,n-1}$}] (m21) at (4,2) {};
	\node[mass,label={[xshift=0.9cm, yshift=0.05cm]$_{m+1,n-1}$}] (m31) at (6,2) {};
	\coordinate (r1) at (8,2);
	
	\coordinate (l2) at (0,4);
	\node[mass,label={[xshift=0.9cm, yshift=0.05cm]$_{m-1,n}$}] (m12) at (2,4) {};
	\node[mass,label={[xshift=0.9cm, yshift=0.05cm]$_{m,n}$}] (m22) at (4,4) {};
	\node[mass,label={[xshift=0.9cm, yshift=0.05cm]$_{m+1,n}$}] (m32) at (6,4) {};
	\coordinate (r2) at (8,4);
	
	\coordinate (l3) at (0,6);
	\node[mass,label={[xshift=0.9cm, yshift=0.05cm]$_{m-1,n+1}$}] (m13) at (2,6) {};
	\node[mass,label={[xshift=0.9cm, yshift=0.05cm]$_{m,n+1}$}] (m23) at (4,6) {};
	\node[mass,label={[xshift=0.9cm, yshift=0.05cm]$_{m+1,n+1}$}] (m33) at (6,6) {};
	\coordinate (r3) at (8,6);

	\draw[spring=4pt] (l1) -- node[above] {} (m11);
	\draw[spring=4pt] (m11) -- node[above] {} (m21);
	\draw[spring=4pt] (m21) -- node[above] {} (m31);
	\draw[spring=4pt] (m31) -- node[above] {} (r1);
	
	\draw[spring=4pt] (l2) -- node[above] {} (m12);
	\draw[spring=4pt] (m12) -- node[above] {} (m22);
	\draw[spring=4pt] (m22) -- node[above] {} (m32);
	\draw[spring=4pt] (m32) -- node[above] {} (r2);
	
	\draw[spring=4pt] (l3) -- node[above] {} (m13);
	\draw[spring=4pt] (m13) -- node[above] {} (m23);
	\draw[spring=4pt] (m23) -- node[above] {} (m33);
	\draw[spring=4pt] (m33) -- node[above] {} (r3);
	
	\draw[spring=4pt] (2,8) -- node[above] {} (m13);
	\draw[spring=4pt] (m13) -- node[above] {} (m12);
	\draw[spring=4pt] (m12) -- node[above] {} (m11);
	\draw[spring=4pt] (m11) -- node[above] {} (2,0);
	
	\draw[spring=4pt] (4,8) -- node[above] {} (m23);
	\draw[spring=4pt] (m23) -- node[above] {} (m22);
	\draw[spring=4pt] (m22) -- node[above] {} (m21);
	\draw[spring=4pt] (m21) -- node[above] {} (4,0);
	
	\draw[spring=4pt] (6,8) -- node[above] {} (m33);
	\draw[spring=4pt] (m33) -- node[above] {} (m32);
	\draw[spring=4pt] (m32) -- node[above] {} (m31);
	\draw[spring=4pt] (m31) -- node[above] {} (6,0);
	
	\end{tikzpicture}
	\caption{A mass--spring system arranged in a 2D square lattice. The masses are fixed at the $(m,n)$ sites with the vertical displacements given by $q_{m,n}(t)$.}
	\label{fig:2Dlat}
\end{figure}
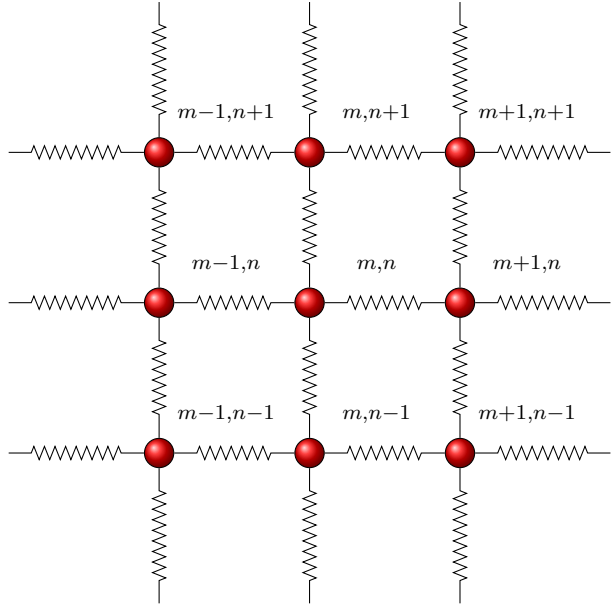

We consider a scalar Fermi--Pasta--Ulam (FPU) system on a square 2D lattice. 
The equations of motion are given by 
\begin{eqnarray}  
\nonumber
\partial_t^2 q_{m,n} & = & W'(q_{m+1,n}-q_{m,n})- W'(q_{m,n}-q_{m-1,n})\\ && +W'(q_{m,n+1}-q_{m,n})- W'(q_{m,n}-q_{m,n-1}), \quad (m,n) \in \Z^2, \label{fpuintro}
\end{eqnarray}
where the scalar variable $ q_{m,n}(t) $ describes a vertical displacement in $ z $-direction of a particle with unit mass located at the $(m,n)$-th site in the $(x,y)$ plane. The interaction potential between a particle and its four neighborhs is described by $W$. Figure \ref{fig:2Dlat} shows the mass--spring system on a square 2D lattice.

The total conserved energy of the FPU system (\ref{fpuintro}) is given by 
\begin{eqnarray}  
\label{energy}
\mathcal{H}(q) = \sum_{(m,n) \in \Z^2}
\frac{1}{2} (\partial_t q_{m,n})^2 + W(q_{m+1,n}-q_{m,n}) + W(q_{m,n+1}-q_{m,n}). \label{fpu-energy}
\end{eqnarray}
For notational simplicity, we assume $ W(u) = \frac{1}{2} u^2 - \frac{1}{3} u^3 $ so that $ W'(u) = u -u^2 $.

It is well-known \cite{KdV-review} that small-amplitude long-scale waves of the FPU system 
in one spatial dimension are described by the Korteweg--de Vries (KdV) equation. 
The KdV equation was first justified with bounds on the approximation error on the unbounded domain in \cite{SW99} and on the periodic domain in \cite{BP,Ponno}. Applications of these methods for other generalized KdV equations  can be found in \cite{DP,GPR,Herr1,Khan}. Properties of solitary waves in the FPU system in one spatial dimension were recently reviewed in \cite{Vainshtein}.

We are interested to justify the Kadomtsev--Petviashvili (KP-II) approximation which describes unidirectional long waves of small amplitude with slowly varying  transverse modulations. The formal derivation of the KP-II equation was reported in \cite{Z} but the rigorous justification was considered to be an 
open problem for some time. Two rigorous results were obtained 
only very recently. The KP-II equation was justified in 
the periodic domain among other integrable normal forms \cite{PP21}. 
By using a more general setting of the vector FPU systems with particles moving in the $(x,y)$ directions, the KP-II equation was justified in the unbounded 2D square lattice 
for the propagation along the axes (as well as for the diagonal propagation 
in the $(x,y)$-plane under additional constraints on the parameters of the lattice) \cite{niki}. 

The KP-II approximation is different from the KdV equation derived 
in the vector FPU systems with geometric nonlinearities, where 
small-amplitude supersonic longitudinal solitary waves may propagate along the horizontal direction  \cite{FM-2003} and along arbitrary directions \cite{CH}. 
Similarly, the liinearized KdV equation was derived for 
the linear propagation of rings in two-dimensional lattices \cite{ST}, 
where the diffraction properties were neglected. 

The purpose of this paper is to improve the justification result 
obtained in \cite{niki} so that it could apply to all directions of 
propagation in the $(x,y)$ plane and without additional restrictions 
on parameters of the lattice. We take the normalized potential $W$ 
and consider the scalar FPU lattice for simplicity, although extensions 
to the vector case with more complicated potentials $W$ are relatively 
straightforward. The main novelty of our approach compared to \cite{niki} is working in Fourier space. Additionally, we have to construct a higher-order approximation involving the linearized KP-II equation in order to handle arbitrary directions of the wave propagation. 

The paper is organized as follows. In Section \ref{sec1} we introduce the strain variables for which the KP-II equations can be derived. By introducing the strain variables the original system is doubled and an additional compatibility condition has to be satisfied by the two strain variables.
The new variables are transformed in Fourier space.

In Section \ref{sec2}  we derive the KP-II equation for the wave propagation 
along the $x$-direction. This simple case was considered in \cite{niki} 
and is used here to highlight analysis from 
the more complicated case of propagation along an arbitrary direction.  
We use smooth solutions of the KP-II equation to construct a suitable approximation of the FPU system satisfying the compatibility condition. The residual terms from the FPU system  with the leading-order approximation 
are estimated, after which the main approximation result is formulated. 

We prove the approximation result in Section \ref{sec4}, where we derive equations 
for the error terms produced by the leading-order approximation 
and handle these terms by using energy estimates and Gronwall's inequality. 
Many terms are miraculously given by the time derivative 
of the energy quantity due to the energy conservation  (\ref{energy}). 

Section \ref{sec3} extends the approximation result for the wave propagation along an arbitrary direction. To keep the residual terms at the same small order, 
we have to use a higher order approximation leading to the system of the KP-II equation and the linearized KP-II equation as approximation equations.  
The requirements on smooth solutions of the KP-II and linearized KP-II equations in Sobolev spaces of higher regularity are stated as assumptions for the approximation result. 

Section \ref{sec-last} discusses the spatial configurations for which these requirements can be satisfied, e.g. 
for transversely independent solutions and for periodic solutions, 
and formulates an open problem for existence of smooth decaying 
solutions in the unbounded domain. \\

{\bf Acknowledgement.} The work of D. E. Pelinovsky is partially supported by the Alexander von Humboldt Foundation as Humboldt Reseach Award.  The work of G. Schneider is partially supported by the Deutsche Forschungsgemeinschaft DFG through the SFB 1173 ''Wave phenomena'' Project-ID 258734477.

\section{Strain variables and the compatibility condition}
\label{sec1}

For further work we introduce 
the following strain variables 
\begin{equation} \label{strain}
u_{m,n} = q_{m+1,n}-q_{m,n}, \qquad v_{m,n} = q_{m,n+1}-q_{m,n}, \qquad 
w_{m,n} = \partial_t q_{m,n}.
\end{equation}
The scalar FPU system (\ref{fpuintro}) can be rewritten in the strain 
variables as the following system
\begin{equation}
\label{fpu-system}
\left\{ \begin{array}{l} 
\partial_t u_{m,n} = w_{m+1,n}-w_{m,n}, \\
\partial_t v_{m,n} = w_{m,n+1}-w_{m,n}, \\
\partial_t w_{m,n} = W'(u_{m,n}) - W'(u_{m-1,n}) + W'(v_{m,n}) - W'(v_{m,n-1}).
\end{array} \right.
\end{equation}
Alternatively, the component $w_{m,n}$ can be eliminated 
and the FPU system (\ref{fpu-system}) can be closed as two 
scalar equations
\begin{eqnarray} 
 \label{umneq} 
\left\{ \begin{array}{l} 
\partial_t^2 u_{m,n} =  W'(u_{m+1,n})- 2 W'(u_{m,n}) + W'(u_{m-1,n})
\\ 
\qquad + W'(v_{m+1,n})- W'(v_{m+1,n-1}) -W'(v_{m,n})+ W'(v_{m,n-1}), 
\\
\partial_t^2 v_{m,n} =  W'(v_{m,n+1})- 2 W'(v_{m,n}) + W'(v_{m,n-1})  
\\ \qquad + W'(u_{m,n+1})- W'(u_{m-1,n+1}) -W'(u_{m,n})+ W'(u_{m-1,n}), 
\end{array} \right.
\end{eqnarray}
where the $ u $- and $ v $-variables satisfy a certain compatibility condition
due to their relation (\ref{strain}) to the $ q $-variable. 

In order to specify the compatibility condition and to develop the justification 
analysis of the KP-II approximation, we will work in Fourier space, 
similar to analysis in \cite{Schn10} of the FPU system on 1D lattice. Therefore, we define
$$
\hat{u}(k,l,t) = \frac{1}{2\pi} \sum_{(m,n) \in \mathbb{Z}^2} u_{m,n} e^{ikm+iln}, \qquad 
u_{m,n} = \frac{1}{2\pi} \iint_{\mathbb{T}^2} \hat{u}(k,l,t) e^{-ikm-iln} dk dl,
$$ 
and similarly for $v_{m,n}$ and $w_{m,n}$, where $\mathbb{T} := [-\pi,\pi)$ equipped with periodic boundary conditions. The first two equations 
of the system (\ref{fpu-system}) shows that 
$$
\partial_t \hat{u} = (e^{-ik} - 1) \hat{w}, \quad \partial_t \hat{v} = (e^{-il} - 1) \hat{w},
$$
which implies that the following compatibility condition is invariant with 
respect to the time evolution of the FPU system (\ref{fpu-system}):
\begin{equation} \label{uvrelation}
(e^{-ik}-1)  \widehat{v}(k,l,t)  =  (e^{-il}-1) \widehat{u}(k,l,t), 
\quad t \geq 0.
\end{equation}

\begin{remark}
{\rm In general, an arbitrary constant can be added to (\ref{uvrelation}). 
We set this constant to $0$ for the class of solutions we are interested in. }
\end{remark}

Next we rewrite system (\ref{umneq}) with $W'(u) = u - u^2$ in the convenient Fourier form. To do so, let us first inspect the linearized system 
\begin{eqnarray*} 
\left\{ \begin{array}{l} 
\partial_t^2 u_{m,n} =  u_{m+1,n}- 2 u_{m,n} + u_{m-1,n} + v_{m+1,n}- v_{m+1,n-1} -v_{m,n}+ v_{m,n-1}, \\
\partial_t^2 v_{m,n} = v_{m,n+1}- 2 v_{m,n} + v_{m,n-1}
+ u_{m,n+1}- u_{m-1,n+1} -u_{m,n}+ u_{m-1,n},
\end{array} \right.
\end{eqnarray*}
which is written in Fourier space as
\begin{eqnarray*} 
	\left\{ \begin{array}{l} 
\partial_t^2 \widehat{u} = (e^{-i k}-2+e^{i k}) \widehat{u}  + (e^{-ik}-1)(1-e^{il})\widehat{v}, \\
\partial_t^2 \widehat{v} = (e^{-i l}-2+e^{i l}) \widehat{v}  + (e^{-il}-1)(1-e^{ik})\widehat{u}, 
\end{array} \right.
\end{eqnarray*}
where we have used
$$ 
e^{-ik}- e^{-ik}e^{il}-1 + e^{il} = e^{-ik}(1-e^{il}) - (1-e^{il})= (e^{-ik}-1)(1-e^{il}).
$$ 
To simplify notation, we define
$$
\omega_k^2 := 2-e^{-i k}-e^{i k}, \qquad 
\omega_l^2 := 2-e^{-i l}-e^{i l}.
$$
Eliminating $\hat{v}$ in the first equation of the linearized system yields the following linear equation 
\begin{equation}
\label{linear-pr}
\partial_t^2 \widehat{u} + (\omega_k^2 + \omega_l^2) \widehat{u} = 0.
\end{equation} 
Extending exactly the same calculations as for the linearized system 
for $W'(u) = u - u^2$, we obtain the following nonlinear system 
in Fourier space given by 
\begin{eqnarray*} 
		\left\{ \begin{array}{l} 
\partial_t^2 \widehat{u} =  -\omega_k^2(\widehat{u} - \widehat{u}*\widehat{u})+ (e^{-ik}-1)(1-e^{il})(\widehat{v}- \widehat{v}*\widehat{v}), \\
\partial_t^2 \widehat{v} =  -\omega_l^2  (\widehat{v} - \widehat{v}*\widehat{v}) + (e^{-il}-1)(1-e^{ik})(\widehat{u} - \widehat{u}*\widehat{u}). 
\end{array} \right.
\end{eqnarray*}
By using the compatibility condition \eqref{uvrelation}, we rewrite this system in the form:
\begin{equation} \label{syst1}
	\left\{ \begin{array}{l} 
\partial_t^2 \widehat{u} = -(\omega_k^2 + \omega_l^2) \widehat{u} + \omega_k^2
(\widehat{u}*\widehat{u}) - (e^{-ik}-1)(1-e^{il})(\widehat{v}*\widehat{v}), 
\\  
\partial_t^2 \widehat{v} = -(\omega_k^2 + \omega_l^2) \widehat{v} + \omega_l^2(\widehat{v}*\widehat{v})- (e^{-il}-1)(1-e^{ik})( \widehat{u}*\widehat{u}). 
\end{array} \right.
\end{equation}
This system in combination with the compatibility condition \eqref{uvrelation} is the starting point for the derivation and justification 
of the KP-II equation. 

\begin{remark}{\rm 
By using the compatibility condition \eqref{uvrelation}, the two equations 
in system \eqref{syst1} can be reduced to a single equation.
However, this single equation contains multipliers which are singular 
with respect to $ k $ and $ l $. Since 
we have to expand these multipliers with respect to $ k $ and $ l $ 
in the long wave limit $(k,l) \to (0,0)$, it is advantageous to work with system
\eqref{syst1} where the multipliers are smooth with respect to $ k $ and $ l $.
}
\end{remark}

\section{Propagation along the $ x $-direction}
\label{sec2}

Here we first derive the KP-II equation for the long modulated waves moving along the $ x $-axis. After deriving the KP-II equation, we discuss how to handle the leading-order approximation in Fourier space and the residual terms of the FPU system. We end this section by formulating the approximation theorem, which will be proven in Section \ref{sec4}.

\subsection{The formal long-wave limit}
\label{sec21}

The long wave limit in physical space corresponds in Fourier space to an 
expansion of system \eqref{syst1} at the wave vector $(k,l) = (0,0)$.
Expansions
$$ 
\omega_k^2 = k^2 - \frac{1}{12} k^4  + \mathcal{O}(k^6), \qquad 
1-e^{il} = -il + \mathcal{O}(l^2),
$$ 
allows us to rewrite system (\ref{syst1}) in physical space formally as
\begin{equation} 
\label{start3}
\partial_t^2 u = \partial_x^2 u  +  \partial_y^2 u  +  \frac{1}{12} \partial_x^4 u  +  \frac{1}{12} \partial_y^4 u 
-  \partial_x^2 (u^2)  - \partial_x  \partial_y (v^2) + \mbox{\rm h.o.t.},
\end{equation}
where $\mbox{\rm h.o.t.}$ stands for the higher-order terms. 
The compatibility condition (\ref{uvrelation}) corresponds in physical space 
to 
\begin{equation}
\label{start2}
\partial_x v + \mbox{\rm h.o.t.} = \partial_y u + \mbox{\rm h.o.t.}.
\end{equation}

The leading-order approximation is given in physical space by 
\begin{equation}
\label{approx1}
u_{m,n}(t) = \varepsilon^2 A(X,Y,T), \quad v_{m,n}(t) = \varepsilon^3 \partial_X^{-1} \partial_Y A(X,Y,T),
\end{equation} 
with
\begin{equation*}
X =  \varepsilon(m - t),  
\quad Y = \varepsilon^2 n,  \quad T = \varepsilon^3 t,
\end{equation*}
where $A$ is a suitable solution to the KP-II equation (\ref{kp2}) below 
for which derivatives of $A$ and $\partial_X^{-1} \partial_Y A$ are controlled in Sobolev spaces of sufficiently high regularity. The compatibility condition (\ref{start2}) rewritten in variables $(X,Y,T)$ is satisfied at the order of $\mathcal{O}(\varepsilon^4)$. Substitution of (\ref{approx1}) into (\ref{start3}) rewritten in variables $(X,Y,T)$ results in the 
following KP-II equation at the order of $\mathcal{O}(\varepsilon^6)$:
\begin{equation}
\label{kp2}
2 \partial_X \partial_T A + \partial_Y^2 A  +  \frac{1}{12} \partial_X^4 A - \partial_X^2 (A^2) = 0.
\end{equation}
In what follows, we replace the formal approximation in physical space 
by the precise approximation in Fourier space.

\subsection{The leading-order approximation in Fourier space}

Our goal is to prove a statement of the following form.
Let $ A $ be a suitable solution of the KP-II equation \eqref{kp2}. 
Then for $ \varepsilon > 0 $ sufficiently small, there are solutions of 
system \eqref{umneq} which remain close to the leading-order 
approximation (\ref{approx1}). 

In order to establish such an  approximation theorem, we have
to estimate the residual terms first, i.e., we have to control the terms which do not cancel after inserting the approximation (\ref{approx1}) into system \eqref{start3} and \eqref{start2}.
In general, these estimates can be obtained by expanding the 
multipliers in Fourier space and by assuming  a certain 
regularity of the solutions of the KP-II equation (\ref{kp2}).
However, additional difficulties occur as we explain below. 

\begin{enumerate}
	\item[\bf (S1)]
	A fundamental difficulty is the fact that the surface of 
	wave frequencies $\omega := \sqrt{\omega_k^2 + \omega_l^2}$	of the linearized problem (\ref{linear-pr}) forms a cone at the wave vector $ (k,l) = (0,0) $. Hence, a Taylor series expansion of the surface is not possible at the tip of the cone.
	A consequence of non-smoothness is the occurrence of the 
	term $ \partial_X^{-1}  \partial_Y^{2} A $ in the 
	evolution problem 
	\begin{equation}
	\label{KP-evolution}
	\partial_T A = -\frac{1}{2} \left( \partial_X^{-1} \partial_Y^2 A  +  \frac{1}{12} \partial_X^3 A - 2 A \partial_X A\right),
	\end{equation}
	which follows from the KP-II equation (\ref{kp2}).

	\item[\bf (S2)] We partially get rid of the first difficulty (S1) by working with the extended system \eqref{syst1}. However, the leading-order approximation for  the extended system \eqref{syst1} has to satisfy the  compatibility condition \eqref{uvrelation} from which the term $\partial_X^{-1} \partial_Y A$ appear in the approximation (\ref{approx1}). 	
	
	\item[\bf (S3)] By looking at the evolution equation (\ref{KP-evolution}), it cannot be expected that the solutions of the KP-II equation are arbitrarily smooth. However, a certain smoothness of solutions is 
	needed for estimating the residual terms.
		
	\item[\bf (S4)]
	Finally, solutions of the FPU system \eqref{umneq} live on $ \Z^2 $, whereas solutions of the KP-II equation (\ref{kp2}) live on $ \R^2 $.  In Fourier space, solutions of system \eqref{syst1} live on $ \mathbb{T}^2 $, whereas solutions of the Fourier-transformed KP-II equation live on $ \R^2 $, 
	$$
	\hat{A}(\xi,\eta,T) = \frac{1}{2\pi} \iint_{\R^2} A(X,Y,T) e^{i \xi X + i \eta Y} d X d Y
	$$
	and
	$$
	A(X,Y,T) = \frac{1}{2\pi} \iint_{\R^2} \hat{A}(\xi,\eta,T) e^{-i \xi X - i \eta Y} d \xi d \eta.
	$$	
\end{enumerate}

To deal with the difficulties in (S1)--(S3), we use the following well-posedness 
result obtained in \cite[Lemma 1]{niki} based on earlier work \cite{GS-KP,Ukai1989}.

\begin{lemma}
	\label{KPIIwellpos2}
	For any $A_0\in H^{s+9}( \R^2)$ such that 
	$\partial_X^{-2}\partial_Y^2 A_0\in H^{s+9}(\R^2)$ 
	and 
	$$
	\partial_X^{-1}\partial_Y^2( \partial_X^{-2}\partial_Y^2 A_0 - A_0^2)\in H^{s+3}( \R^2)
	$$ 
	with fixed $s \geq 0$, there exists $\tau_0>0$ such that the Cauchy problem \eqref{KP-evolution} admits a unique solution
	$$
	A \in C^0([-\tau_0,\tau_0],H^{s+9})\cap C^1([-\tau_0,\tau_0],H^{s+6})
	\cap C^2([-\tau_0,\tau_0],H^{s+3})
	$$
	such that 
	$$ 
	\partial_X^{-1}\partial_Y A \in C^0([-\tau_0,\tau_0],H^{s+8}) \cap
	C^1([-\tau_0,\tau_0],H^{s+5}) \cap
	C^2([-\tau_0,\tau_0],H^{s+2}) 
	$$
	and
	$$
	\partial_X^{-2}\partial_Y^2  A \in C^0([-\tau_0,\tau_0],H^{s+6})  \cap C^1([-\tau_0,\tau_0],H^{s+3}).
	$$
\end{lemma}

To deal with the difficulty in (S4), we can use the following approximation result from \cite[Lemma 2]{niki}, which was obtained based on 
previous estimates in one dimension in \cite[Lemma 3.9]{SW99} and \cite[Lemma 5.1]{DP}. 

\begin{lemma} 
	\label{multiplier}
	Let $ A \in C^0([-\tau_0,\tau_0],H^s(\R^2))$ with $s > 1$ and 
	$a_{m,n}(t) := A(\varepsilon (m-t), \varepsilon^2 n, \varepsilon^3 t)$ 
	for $(m,n) \in \Z^2$. Then there exists a constant $ C_s > 0 $ such that for all $ \varepsilon \in (0,1] $ we have 
	$$ 
	\| a(t)  \|_{\ell^2(\Z^2)} \leq C_s \varepsilon^{-\frac{3}{2}} \| A(\cdot,\cdot,\varepsilon^3 t) \|_{H^s(\mathbb{R}^2)}, \quad \forall t \in [-\varepsilon^{-3} \tau_0,\varepsilon^{-3} \tau_0].
	$$ 
	Consequently, in Fourier space, we have 
$$ 
\| \hat{a}(\cdot,\cdot,t)  \|_{L^2(\mathbb{T}^2)} \leq C_s \varepsilon^{-\frac{3}{2}} \| \hat{A}(\cdot,\cdot,
\varepsilon^3 t) \|_{L^{2,s}(\mathbb{R}^2)}, \quad \forall t \in [-\varepsilon^{-3} \tau_0,\varepsilon^{-3} \tau_0],
$$ 
where $\| \hat{A} \|_{L^{2,s}(\R^2)} := \| <\cdot>^s \hat{A}\|_{L^2(\R^2)}$ with $<x> := \sqrt{1 + |x|^2}$.
\end{lemma}

\begin{remark}
	{\rm The proof of Lemma \ref{multiplier} is well-known in the existing literature, cf. the book \cite{SU17}. In Fourier space we lose $ \varepsilon^{-3} $ due to the Fourier transform of the scaled variables and gain $ \varepsilon^{3/2} $ due to the scaling of the $ L^2 $-norm. This coincides with the estimates in physical space, where we lose a factor $ \varepsilon^{-3/2} $ due to scaling. The bounds in Fourier and physical space agree to each other since Fourier transform is an isomorphism in $ L^2 $ spaces.
	}
\end{remark}

Finally, we make precise the leading-order approximation in Fourier space, 
which we denote as $(\hat{u},\hat{v}) = \varepsilon^2 (\hat{\psi}_u,\hat{\psi}_v)$. Let $\hat{A}$ be the Fourier transform 
of a smooth solution $A$ of the KP-II equation (\ref{kp2}) in Lemma \ref{KPIIwellpos2}. Let $\chi_{\mathbb{T}^2}$ be the characterstic function 
on $\mathbb{R}^2$ such that $\chi_{\mathbb{T}^2}(k,l) = 0$ for $(k,l) \notin \mathbb{T}^2$. Then, $\hat{\psi}_u$ is defined by  
\begin{equation} 
\label{scali}
\hat{\psi}_u(k,l,t) = \varepsilon^{-3} e^{ikt} \chi_{\mathbb{T}^2}(k,l) \widehat{A} ( \varepsilon^{-1} k, \varepsilon^{-2} l,T),
\end{equation}
whereas $\hat{\psi}_v$ is obtained from the compatibility relation
\begin{equation}
\label{compt-leading}
(e^{-ik}-1)  \widehat{\psi}_v(k,l,t)  =  (e^{-il}-1) \widehat{\psi_u}(k,l,t).
\end{equation}
Expanding as $(k,l) \to (0,0)$ yields 
\begin{equation} 
		\label{scal-v}
\widehat{\psi}_v(k,l,t) = k^{-1} l \left[ 1 + \mathcal{O}(|k| + |l|) \right] \widehat{\psi_u}(k,l,t).
\end{equation}
It is clear that the inverse Fourier transform of (\ref{scali}) and (\ref{scal-v}) does not recover the approximation (\ref{approx1}) in physical space because of the approximation errors. 
However, the following lemma controls the difference between the approximations 
in Fourier and physical spaces. The lemma is based on \cite[Lemma 3.8]{SW99} 
in one dimension and can be proven similarly to Lemma \ref{multiplier} proven in \cite{niki}.

\begin{lemma} 
	\label{lem-approx}
	Let $ A \in C^0([-\tau_0,\tau_0],H^s(\R^2))$ with $s > 1$, 
	$a_{m,n}(t) := A(\varepsilon (m-t), \varepsilon^2 n, \varepsilon^3 t)$ and $\hat{\psi}_u$ be given by (\ref{scali}). Then, there exists a constant $ C_s > 0 $ such that for all $ \varepsilon \in (0,1] $ we have 
	$$ 
	\| \psi_u(t) - a(t) \|_{\ell^2(\Z^2)} \leq C_s \varepsilon^{s-\frac{3}{2}} \| A(\cdot,\cdot,\varepsilon^3 t) \|_{H^s(\mathbb{R}^2)}, \quad \forall t \in [-\varepsilon^{-3} \tau_0,\varepsilon^{-3} \tau_0].
	$$ 
\end{lemma}

\begin{remark}
{\rm In view of Lemma \ref{multiplier}, the difference between the approximation (\ref{approx1}) 
in physical space and $(\varepsilon^2 \psi_u,\varepsilon^2 \psi_v)$ with $\psi_u$ and $\psi_v$ given by 
the inverse Fourier transform of the approximation (\ref{scali}) and (\ref{scal-v}) is small.}
\end{remark}

\subsection{Estimates for the residual}

The residuals contain the terms which do not cancel after 
inserting the leading-order approximation (\ref{scali}) and (\ref{scal-v}) into system \eqref{syst1}:
\begin{eqnarray*}
\left\{ \begin{array}{l}
	\widehat{\textrm{Res}}_u(u,v) :=
	-\partial_t^2 \widehat{u}(\ell,t) -(\omega_k^2(k) + \omega_l^2(l)) \widehat{u} (\ell,t)+ \omega_k^2(k)
	(\widehat{u}*\widehat{u})(\ell,t) \\ 
	\qquad \qquad \qquad - (e^{-ik}-1)(1-e^{il})(\widehat{v}*\widehat{v})(\ell,t), 
	\\
	\widehat{\textrm{Res}}_v(u,v) :=
	-\partial_t^2 \widehat{v}(\ell,t)  -(\omega_k^2(k) + \omega_l^2(l)) \widehat{v} (\ell,t)+ \omega_l^2(l)(\widehat{v}*\widehat{v}) (\ell,t)
	\\ \qquad \qquad \qquad - (e^{-il}-1)(1-e^{ik})( \widehat{u}*\widehat{u})(\ell,t).
	\end{array} \right.
\end{eqnarray*}

\begin{remark}
	{\rm 
		The application of Lemmas \ref{multiplier} and \ref{lem-approx} transfers the pure counting of powers 
		of $ \varepsilon $ into rigorous estimates. Since this is well documented in the 
		existing literature, cf. \cite{SU17}, we refrain from many details. 
	}
\end{remark}

In physical space, it follows formally from (\ref{start3})  and (\ref{approx1}) that 
\begin{eqnarray*}
{\textrm{Res}}_u(\varepsilon^2 \psi_u, \varepsilon^2 \psi_v)  =
	\mathcal{O}( \varepsilon^8 \partial_T^2 A , 
	\varepsilon^8  \partial_X^6 A,
	\varepsilon^{10}  \partial_Y^4 A , \varepsilon^8   \partial_X^4 (A^2) ,   \varepsilon^9   \partial_X \partial_Y (\partial_X^{-1} \partial_Y A)^2)
\end{eqnarray*}
and similarly 
\begin{eqnarray*}
	{\textrm{Res}}_v(\varepsilon^2 \psi_u, \varepsilon^2 \psi_v) =
	\mathcal{O}( \varepsilon^9 \partial_X^{-1} \partial_Y \partial_T^2 A , 
	\varepsilon^9  \partial_X^5 \partial_Y A,
	\varepsilon^{11}  \partial_X^{-1} \partial_Y^5 A , \varepsilon^9   \partial_X^3 \partial_Y (A^2) ,   \varepsilon^{10}  \partial_Y^2 (\partial_X^{-1} \partial_Y A)^2)
\end{eqnarray*}

\begin{remark}
	{\rm The notations are to be understood in the following sense. 
		The term $\mathcal{O}(	\varepsilon^8 \partial_X^6 A)$ means that in Fourier space the scaled $ \widehat{A} $ is multiplied by a function $ f = f(k,l) $ satifying $ |f(k,l) | \leq C 	\varepsilon^8 |k|^6 $, where 
		possibly different constants are denoted with the same symbol $ C $ if they can be chosen independent of the small perturbation parameter $ 0 < \varepsilon \ll 1 $. This estimate is only relevant for small $ k $ and $ l $ since our equations in Fourier space are posed on the bounded domain $ \mathbb{T}^2 $.
}
\end{remark}

The residual terms are controlled by the local well-posedness theory of Lemma \ref{KPIIwellpos2} and by the error bounds of Lemma \ref{multiplier} when $A$ is a smooth solution of the KP-II equation (\ref{KP-evolution}) for any $s \geq 0$. Since we apply $\omega_k^{-1}$ to $\widehat{\textrm{Res}}_u(\varepsilon^2 \psi_u, \varepsilon^2 \psi_v) $  and $\omega_l^{-1}$ to $\widehat{\textrm{Res}}_v(\varepsilon^2 \psi_u, \varepsilon^2 \psi_v) $, both terms in the physical space yield 
\begin{eqnarray*}
\mathcal{O}( \varepsilon^7 \partial_X^{-1} \partial_T^2 A , 
\varepsilon^7  \partial_X^5 A,
\varepsilon^{9}  \partial_X^{-1} \partial_Y^4 A , \varepsilon^7   \partial_X^3 (A^2) ,   \varepsilon^8  \partial_Y (\partial_X^{-1} \partial_Y A)^2)
\end{eqnarray*}
Hence, we lose a factor $ \varepsilon^{-1} $ due to the long wave character of $ \psi_u $ and $ \psi_v $. By taking the $ L^2 $-norm we lose another factor $ \varepsilon^{-3/2} $ due to the involved scalings and the scaling 
properties of the $ L^2 $-norm.
Therefore, it follows from the formal order $\mathcal{O}(\varepsilon^8)$ of 
truncation that we have in the end
\begin{equation} \label{curry}
\|  \omega_k^{-1}
\widehat{\textrm{Res}}_u(\varepsilon^2 \psi_u, \varepsilon^2 \psi_v) \|_{L^2} +  \| \omega_l^{-1}
\widehat{\textrm{Res}}_v(\varepsilon^2 \psi_u, \varepsilon^2 \psi_v) \|_{L^2} = \mathcal{O}(\varepsilon^{\frac{11}{2}}) .
\end{equation}
This means that on the long $ \mathcal{O}(1/\varepsilon^{3}) $-time scale
we can choose the error to scale with a factor $ \varepsilon^{\beta} $ with  
$ \beta = \frac{5}{2} $. The precise count of the residual terms with the same bound (\ref{curry}) was obtained in \cite[Lemma 3]{niki}.

\begin{remark}{\rm 
		The  formal order $ \mathcal{O}(\varepsilon^8) $ is sufficient 
		in the subsequent energy estimates of Section \ref{sec4} 
		since many terms can be written as time-derivatives
		which allows us to include these terms in the chosen energy and allow for  
		$ \beta = \frac{5}{2} $.  If the corresponding terms could not have  been written 
		as time-derivatives, then it would be necessary to obtain 
residual terms of  formal order of $ \mathcal{O}(\varepsilon^9) $, for which 
		we would have to construct a higher order approximation to 
		the leading-order approximation $(u,v) = (\varepsilon^2 \psi_u,\varepsilon^2 \psi_v)$.
	}
\end{remark}

Additional residual terms arise from 
the KP-II equation (\ref{kp2}) rewritten in Fourier space and truncated 
on $\mathbb{T}$. The approximation error obtained from this truncation 
only appears for the nonlinear (quadratic) term of the KP-II equation. 
Since $\mathbb{T}^2$ is compact, it suffices to control this error by considering the convolution term for $A^2$ in Fourier space:
\begin{align*}
&\chi_{\mathbb{T}^2} (\hat{A} \ast \hat{A}) - \chi_{\mathbb{T}^2} (\chi_{\mathbb{T}^2} \hat{A} \ast \chi_{\mathbb{T}^2} \hat{A}) \\
&=  \chi_{\mathbb{T}^2} ((\hat{A} - \chi_{\mathbb{T}^2} \hat{A}) \ast (\hat{A} - \chi_{\mathbb{T}^2} \hat{A})) + 2 \chi_{\mathbb{T}^2} (\chi_{\mathbb{T}^2} \hat{A} \ast (\hat{A} - \chi_{\mathbb{T}^2} \hat{A})).
\end{align*}
Each term in the right-hand side is controlled by the application 
of Lemma \ref{lem-approx} and the residual error in Fourier space 
is smaller than the bound (\ref{curry}) for any $s > 1$.

\subsection{The approximation result}

By using the energy estimates for the approximation error, 
we will prove  in Section \ref{sec4} the following main result for the horizontal propagation in the two-dimensional square lattice. 

\begin{theorem}
	\label{th-hor}
There exist  $ C_0 $ and $ \varepsilon_0 > 0 $ such that for all $ \varepsilon  \in (0, \varepsilon_0) $ the following holds. Let $ A \in 
	C([0,\tau_0], H^{s+9})  $ be a solution of the KP-II equation \eqref{kp2} given by 
	Lemma \ref{KPIIwellpos2} with fixed $s \geq 0$. Then there exist solutions $ (u,v) $ of system \eqref{umneq} with 
	$$
	\sup_{t \in [0,\varepsilon^{-3} \tau_0]} \| u(t) - \varepsilon^2 \psi_u(t) \|_{\ell^2(\Z^2)} + \| v(t) - \varepsilon^2 \psi_v(t) \|_{\ell^2(\Z^2)}  \leq C_0 \varepsilon^{\frac{5}{2}},
	$$
	where $(\psi_u,\psi_v)$ are given by the inverse Fourier transform 
	of (\ref{scali}) and (\ref{scal-v}).
\end{theorem}

\begin{remark}
	{\rm
		The proof of the approximation result of Theorem \ref{th-hor} is a nontrivial task. The KP-II approximation and 
		the associated solution  are of order $ \mathcal{O}(\varepsilon^2) $ for $ \varepsilon \to 0 $. Therefore, a simple application of Gronwall's inequality would only provide the boundedness of the solutions on an $ \mathcal{O}(\varepsilon^{-2}) $-time scale, but not on the natural $ \mathcal{O}(\varepsilon^{-3}) $-time scale of the KP approximation.
		There exist many counterexamples where formally derived amplitude equations make wrong predictions  about the dynamics of original systems on the natural time scale of the amplitude equations, cf.~\cite{Schn96MN} and recent results in \cite{BSSZ20,HS20,SSZ15}. 
	}
\end{remark}

\section{Energy estimates for the approximation error}
\label{sec4}

The approximation error is defined by 
\begin{equation}
\label{error-terms}
\varepsilon^{\beta} (\hat{R}_u,\hat{R}_v) := (\hat{u},\hat{v})- (\varepsilon^{2} \widehat{\psi}_u,\varepsilon^{2} \widehat{\psi}_v) 
\end{equation}
with $\beta$ being suitably chosen as $\beta = \frac{5}{2}$, the approximation $ (\varepsilon^{2} \widehat{\psi}_u,\varepsilon^{2} \widehat{\psi}_v)  $ satisfying the compatibility condition (\ref{compt-leading}), and the error 
terms $(\hat{R}_u,\hat{R}_v)$ satisfying the compatibility condition 
\begin{equation}
\label{compt-error}
(e^{-ik}-1)  \widehat{R}_v(k,l,t)  =  (e^{-il}-1) \widehat{R}_u(k,l,t).
\end{equation}
The error terms satisfy equations of motion given by 
\begin{align} 
\nonumber
\partial_t^2 \widehat{R}_u &=-(\omega_k^2 + \omega_l^2)  \widehat{R}_u + 2 \varepsilon^{2} \omega_k^2 
(\widehat{\psi}_u*\widehat{R}_u) +  \varepsilon^{\beta}  \omega_k^2 
(\widehat{R}_u*\widehat{R}_u) \\ 
\nonumber
& \qquad  - 2 \varepsilon^{2} (e^{-ik}-1)(1-e^{il})(\widehat{\psi}_v*\widehat{R}_v)-\varepsilon^{\beta}   (e^{-ik}-1)(1-e^{il})(\widehat{R}_v*\widehat{R}_v) \\ 
\label{error-R-u}
& \qquad+  \varepsilon^{- \beta} \widehat{\textrm{Res}}_u(\varepsilon^{2} \psi_u, \varepsilon^{2} \psi_v), \\
\nonumber
\partial_t^2 \widehat{R}_v & = -(\omega_k^2 + \omega_l^2) \widehat{R}_v + 2 \varepsilon^{2} \omega_l^2(\widehat{\psi}_v*\widehat{R}_v) + 
\varepsilon^{\beta} \omega_l^2(\widehat{R}_v*\widehat{R}_v) \\ 
\nonumber
& \qquad- 2 \varepsilon^{2} (e^{-il}-1)(1-e^{ik})( \widehat{\psi}_u*\widehat{R}_u) -   \varepsilon^{\beta}  (e^{-il}-1)(1-e^{ik})( \widehat{R}_u*\widehat{R}_u)
\\ 
\label{error-R-v}
&\qquad +  \varepsilon^{- \beta} \widehat{\textrm{Res}}_v(\varepsilon^{2} \psi_u, \varepsilon^{2} \psi_v).
\end{align}

To progress further, we recall the conserved energy (\ref{energy}) of the FPU system (\ref{fpuintro}), which suggests that the energy for the error terms 
in physical space can be defined in the form
$$ 
E = E_0 +  \varepsilon^{2}  E_1 + \varepsilon^{\beta}  E_2
$$
with
\begin{equation}
\label{E-0}
E_0 = \| R_w \|^2_{\ell^2} + \| R_u \|^2_{\ell^2} + \| R_v \|^2_{\ell^2},
\end{equation} 
\begin{equation}
\label{E-1}
E_1 = -2 \sum_{(m,n) \in \Z^2} (\psi_u)_{m,n} (R_u^2)_{m,n} + (\psi_v)_{m,n} (R_v^2)_{m,n}, 
\end{equation}
and
\begin{equation}
\label{E-2}
E_2 = -\frac{2}{3}\sum_{(m,n) \in \Z^2} (R_u^3)_{m,n} + (R_v^3)_{m,n},
\end{equation}
where $R_w$ is the error term for the third strain variable in (\ref{strain}) 
and the total energy (\ref{energy}) is is multiplied by a factor of $2$ for convenience. We define the $ L^2 $-scalar product $ (\cdot,\cdot) $ by
$$
(\widehat{f},\widehat{g}) = \int_{\mathbb{T}^2} \overline{\widehat{f}(\ell)} \widehat{g}(\ell) d\ell,
$$
where $\ell := (k,l)$. By using Parseval's equality, we have  
$ \sum\limits_{(m,n)\in \Z^2} f_{m,n} g_{m,n} = (\hat{f},\hat{g}) $.
By using Fourier transform, the first two linear equations 
in system (\ref{fpu-system}), and the compatibility relation (\ref{compt-error}), we can rewrite the leading-order energy 
in the equivalent form
\begin{align}
\label{E-0-fourier}
E_0 = \frac12 \| \omega_k^{-1} \partial_t \widehat{R}_u \|_{L^2}^2 
+ \frac12 \| \omega_l^{-1} \partial_t \widehat{R}_v \|_{L^2}^2 
+ \frac12 \| \omega_k^{-1} \omega  \widehat{R}_u \|_{L^2}^2 
+ \frac12 \| \omega_l^{-1} \omega  \widehat{R}_v \|_{L^2}^2,
\end{align}
where $\omega := \sqrt{\omega_k^2 + \omega_l^2}$. Indeed, the first term in (\ref{E-0}) after the Fourier transform is split symmetrically 
by using the first two linear equations in system (\ref{fpu-system}) as
\begin{align*}
\| \widehat{R}_w \|_{L^2}^2 &= \frac12 \| \omega_k^{-1} \partial_t \widehat{R}_u \|_{L^2}^2 + \frac12 \| \omega_l^{-1} \partial_t \widehat{R}_v \|_{L^2}^2,
\end{align*}
whereas the second and third terms in (\ref{E-0}) after the Fourier transform 
are rewritten as
$$
\| \widehat{R}_u \|_{L^2}^2  + \| \widehat{R}_v \|_{L^2}^2 = \frac12 \| \omega_k^{-1} \omega  \widehat{R}_u \|_{L^2}^2 
+ \frac12 \| \omega_l^{-1} \omega  \widehat{R}_v \|_{L^2}^2,
$$
since the compatibility relation (\ref{compt-error}) suggests that 
$$
\omega^2 (\omega_l^2 |\widehat{R}_u |^2 + \omega_k^2 |\widehat{R}_v |^2) = 
2 \omega_k^2 \omega_l^2 (\widehat{R}_u |^2 + |\widehat{R}_v |^2).
$$
Similarly, we rewrite $E_1$ and $E_2$ after Fourier transform as  
\begin{equation}
\label{E-1-fourier}
E_1 := - 2   \int \overline{\widehat{R}_u(\ell)}
\widehat{\psi}_u(\ell-\ell_1)  \widehat{R}_u(\ell_1) d \ell_1 d \ell 
- 2 \int \overline{\widehat{R}_v(\ell)}
\widehat{\psi}_v(\ell-\ell_1)   \widehat{R}_v(\ell_1) d \ell_1 d \ell
\end{equation}
and
\begin{equation}
\label{E-2-fourier}
E_2 =  -\frac{2}{3} \int \overline{  \widehat{R}_u(\ell)}
	\widehat{R}_u(\ell-\ell_1)  \widehat{R}_u(\ell_1) d \ell_1 d \ell 
-\frac{2}{3} \int \overline{   \widehat{R}_v(\ell)}
	\widehat{R}_v(\ell-\ell_1)   \widehat{R}_v(\ell_1) d \ell_1 d \ell .
\end{equation}  

The leading-order energy $E_0$ in (\ref{E-0-fourier}) 
suggests that the energy estimates for the system (\ref{error-R-u}) and (\ref{error-R-v}) are obtained by multiplying 
the first equation by the weighted  time derivative $ \omega_k^{-2}  \overline{\partial_t \widehat{R}_u}  $ and the second equation by  the weighted  time derivative $  \omega_l^{-2} \overline{\partial_t \widehat{R}_v} $, after which we add the two equations 
and integrate in $\ell = (k,l)$. This procedure gives us the following energy balance equation 
\begin{equation}
\label{energy-balance} 
\frac12
\partial_t  \| \omega_k^{-1}  \partial_t \widehat{R}_u \|_{L^2}^2 + \frac12 \partial_t \| \omega_l^{-1}  \partial_t \widehat{R}_v \|_{L^2}^2
 =  {\rm Re}(s_1 + s_2 + \ldots  + s_{12}),
\end{equation}
with 
\begin{align*} 
s_1 & = - \int \overline{\partial_t \widehat{R}_u(\ell)}  \omega_k^{-2} \omega^{2} \widehat{R}_u(\ell) d\ell,\\
s_2 & = 2 \varepsilon^{2}  \int \overline{\partial_t \widehat{R}_u(\ell)}
\widehat{\psi}_u(\ell-\ell_1)\widehat{R}_u(\ell_1) d \ell_1 d \ell,\\
s_3 & = \varepsilon^{\beta} \int \overline{\partial_t \widehat{R}_u(\ell)}
\widehat{R}_u(\ell-\ell_1)\widehat{R}_u(\ell_1) d \ell_1 d \ell,\\
s_4 & = -   2 \varepsilon^{2}  \int \overline{\partial_t \widehat{R}_u(\ell)}
(e^{-ik}-1)(1-e^{il}) \omega_k^{-2} \widehat{\psi}_v(\ell-\ell_1)\widehat{R}_v(\ell_1) d \ell_1 d \ell,\\
 s_5 & = -  \varepsilon^{\beta} \int \overline{\partial_t \widehat{R}_u(\ell)}
  (e^{-ik}-1)(1-e^{il}) \omega_k^{-2} \widehat{R}_v(\ell-\ell_1)\widehat{R}_v(\ell_1) d \ell_1 d \ell,\\
s_6 & =  \varepsilon^{-\beta} \int \overline{\partial_t \widehat{R}_u(\ell)} \omega_k^{-2}
\widehat{\textrm{Res}}_u(\varepsilon^{2} \psi_u, \varepsilon^{2} \psi_v)(\ell) d \ell, \\
 s_7 & = -\int \overline{\partial_t \widehat{R}_v(\ell)} \omega_l^{-2} \omega^{2} \widehat{R}_v (\ell) d\ell, \\
s_8 & =  2 \varepsilon^{2} \int \overline{\partial_t \widehat{R}_v(\ell)}
\widehat{\psi}_v(\ell-\ell_1)\widehat{R}_v(\ell_1) d \ell_1 d \ell, \\
s_9 & = \varepsilon^{\beta} \int \overline{\partial_t \widehat{R}_v(\ell)}
\widehat{R}_v(\ell-\ell_1)\widehat{R}_v(\ell_1) d \ell_1 d \ell, \\
 s_{10} & = - 2 \varepsilon^{2}\int \overline{\partial_t \widehat{R}_v(\ell)}
(e^{-il}-1)(1-e^{ik})  \omega_l^{-2} \widehat{\psi}_u(\ell-\ell_1)\widehat{R}_u(\ell_1) d \ell_1 d \ell,
\end{align*}
\begin{align*} 
 s_{11} & = -   \varepsilon^{\beta} \int \overline{\partial_t \widehat{R}_v(\ell)}
 (e^{-il}-1)(1-e^{ik}) \omega_l^{-2} \widehat{R}_u(\ell-\ell_1)\widehat{R}_u(\ell_1) d \ell_1 d \ell, \\
 s_{12} & =   \varepsilon^{- \beta}\int \overline{\partial_t \widehat{R}_v(\ell)}
  \omega_l^{-2} \widehat{\textrm{Res}}_v(\varepsilon^{2} \psi_u, \varepsilon^{2} \psi_v)(\ell) d \ell.
 \end{align*}
We can now deal with different terms of the energy balance equation (\ref{energy-balance}) as follows.\\

\noindent
  i) We rewrite $ s_1 $ and $ s_7 $ as
 $$
 s_1 = - \frac12 \partial_t \int \overline{\widehat{R}_u(\ell)}  \omega_k^{-2}(k) \omega^{2}(\ell)
\widehat{R}_u(\ell) d\ell = - \frac12 \partial_t  \| \omega_k^{-1}  \omega \widehat{R}_u \|_{L^2}^2
  $$
 and  
 $$
 s_7 = - \frac12 \partial_t \int \overline{\widehat{R}_v(\ell)}  \omega_l^{-2}(l) \omega^{2}(\ell)
 \widehat{R}_v (\ell) d\ell - \frac12 \partial_t  \| \omega_l^{-1}  \omega \widehat{R}_v \|_{L^2}^2.
  $$
These terms give the time derivative of the third and fourth terms in 
the leading-order energy $E_0$ in (\ref{E-0-fourier}).  \\

\noindent 
  ii) We  rewrite $ s_2 $ and $ s_8 $ as 
\begin{eqnarray*}
 s_2 =   \varepsilon^{2}  \partial_t \int \overline{ \widehat{R}_u(\ell)}
\widehat{\psi}_u(\ell-\ell_1)  \widehat{R}_u(\ell_1) d \ell_1 d \ell - \varepsilon^{2} \int \overline{\widehat{R}_u(\ell)}
\partial_t \widehat{\psi}_u(\ell-\ell_1)  \widehat{R}_u(\ell_1) d \ell_1 d \ell,
\end{eqnarray*}
and
\begin{eqnarray*}
	s_8 =  \varepsilon^{2}  \partial_t \int \overline{  \widehat{R}_v(\ell)}
	\widehat{\psi}_v(\ell-\ell_1)  \widehat{R}_v(\ell_1) d \ell_1 d \ell - \varepsilon^{2} \int \overline{    \widehat{R}_v(\ell)}
	\partial_t \widehat{\psi}_v(\ell-\ell_1) \widehat{R}_v(\ell_1) d \ell_1 d \ell,
\end{eqnarray*}
The first terms in $s_2$ and $s_8$ define one half of the $\varepsilon^2$ correction $E_1$ to the leading-order energy $E_0$ given by (\ref{E-1-fourier}). 
The other half will come from the terms $s_4$ and $s_{10}$. The second term in $s_2$ is estimated as follows:
\begin{align*} 
\varepsilon^{2} \left| \int \overline{\widehat{R}_u(\ell)}
\partial_t \widehat{\psi}_u(\ell-\ell_1)  \widehat{R}_u(\ell_1) d \ell_1 d \ell \right| &\leq \| \widehat{R}_u \|_{L^2} \| \partial_t \widehat{\psi}_u \ast \widehat{R}_u \|_{L^2} \\
& \leq \| \partial_t \widehat{\psi}_u \|_{L^1} \| \widehat{R}_u \|^2_{L^2}, 
\end{align*}
due to the Cauchy-Schwarz and Young's inequality. 
Since 
$$
\partial_t \widehat{\psi}_u(k,l,t) = i k \varepsilon^{-3} e^{ikt} \chi_{\mathbb{T}^2}(k,l) \widehat{A} ( \varepsilon^{-1} k, \varepsilon^{-2} l,T) 
+ e^{ikt} \chi_{\mathbb{T}^2}(k,l) \partial_T \widehat{A} ( \varepsilon^{-1} k, \varepsilon^{-2} l,T),
$$
we obtain 
$$
\| \partial_t \widehat{\psi}_u \|_{L^1} \leq \varepsilon \| |\cdot| \hat{A}(\cdot,\cdot,T) \|_{L^1} + \varepsilon^3 \| \partial_T \hat{A}(\cdot,\cdot,T) \|_{L^1}.
$$
If $A$ is controlled in Sobolev spaces of high regularity with the help of Lemma \ref{KPIIwellpos2}, then 
\begin{eqnarray*}
\varepsilon^{2}  \left| \int \overline{\widehat{R}_u(\ell)}
\partial_t \widehat{\psi}_u(\ell-\ell_1)  \widehat{R}_u(\ell_1) d \ell_1 d \ell  \right| \leq C \varepsilon^{3} \| \widehat{R}_u \|^2_{L^2}.
\end{eqnarray*}
Similarly, we obtain from (\ref{scal-v}) that 
 \begin{eqnarray*}
\varepsilon^{2}  \left| \int \overline{    \widehat{R}_v(\ell)}
\partial_t \widehat{\psi}_v(\ell-\ell_1) \widehat{R}_v(\ell_1) d \ell_1 d \ell  \right| \leq C \varepsilon^4 \| \widehat{R}_v \|^2_{L^2},
\end{eqnarray*}
where the additional power in $\varepsilon^4$ compared to $\varepsilon^3$ is 
explained by additional power of $\varepsilon$ between $\psi_v$ and $\psi_u$, 
cf. (\ref{approx1}).\\
 
 \noindent
iii) We rewrite $ s_4 $  as
\begin{align*}
s_4 &= 2 \varepsilon^{2} \int \overline{ (e^{-il}-1)\omega_k^{-1} \partial_t \widehat{R}_u(\ell)} (e^{-ik}-1)\omega_k^{-1} \widehat{\psi}_v(\ell-\ell_1)\widehat{R}_v(\ell_1) d \ell_1 d \ell \\
&=    2 \varepsilon^{2} \int \overline{ (e^{-ik}-1) \omega_k^{-1} \partial_t \widehat{R}_v(\ell)}
(e^{-ik}-1)\omega_k^{-1} \widehat{\psi}_v(\ell-\ell_1)\widehat{R}_v(\ell_1) d \ell_1 d \ell \\
&=   2 \varepsilon^{2} \int \overline{  \partial_t \widehat{R}_v(\ell)}
\widehat{\psi}_v(\ell-\ell_1)\widehat{R}_v(\ell_1) d \ell_1 d \ell \\
&=    \varepsilon^{2}  \partial_t \int \overline{   \widehat{R}_v(\ell)}
\widehat{\psi}_v(\ell-\ell_1) \widehat{R}_v(\ell_1) d \ell_1 d \ell -
\varepsilon^{2} \int \overline{   \widehat{R}_v(\ell)}
\partial_t  \widehat{\psi}_v(\ell-\ell_1) \widehat{R}_v(\ell_1) d \ell_1 d \ell,
\end{align*}  
and similarly $ s_{10} $ as
\begin{align*}
s_{10} = \varepsilon^{2}  \partial_t \int  \overline{ \widehat{R}_u(\ell)}
	\widehat{\psi}_u(\ell-\ell_1)  \widehat{R}_u(\ell_1) d \ell_1 d \ell -  \varepsilon^{2} \int \overline{   \widehat{R}_u(\ell)}
	\partial_t  \widehat{\psi}_u(\ell-\ell_1)  \widehat{R}_u(\ell_1) d \ell_1 d \ell.
\end{align*}
Since $s_4 + s_{10} = s_2 + s_8$, the first terms in $s_4$ and $s_{10}$ define the other half of $\varepsilon^2 E_1$, where $E_1$ is given by (\ref{E-1-fourier}), where the second terms in $s_4$ and $s_{10}$ have been 
estimated in (ii). \\

\noindent
iv) We  rewrite $ s_3 $ and $ s_9 $ as 
\begin{eqnarray*}
 s_3 = \frac13 \varepsilon^{\beta}  \partial_t \int \overline{ \widehat{R}_u(\ell)}
\widehat{R}_u(\ell-\ell_1)\widehat{R}_u(\ell_1) d \ell_1 d \ell 
\end{eqnarray*}
and 
\begin{eqnarray*}
 s_9 =   \frac13  \varepsilon^{\beta}    \partial_t \int \overline{\widehat{R}_v(\ell)}
\widehat{R}_v(\ell-\ell_1)\widehat{R}_v(\ell_1) d \ell_1 d \ell .
\end{eqnarray*}
These terms define one half of the $\varepsilon^{\beta}$ correction $E_2$ to the leading-order energy $E_0$ given by (\ref{E-2-fourier}). The other half 
will come from the terms $s_5$ and $s_{11}$.\\

\noindent
v) We rewrite $ s_5 $  as
\begin{align*}
s_5 &= 2 \varepsilon^{\beta} \int \overline{ (e^{-il}-1)\omega_k^{-1} \partial_t \widehat{R}_u(\ell)}
(e^{-ik}-1)\omega_k^{-1} \widehat{R}_v(\ell-\ell_1)\widehat{R}_v(\ell_1) d \ell_1 d \ell \\
&=   2 \varepsilon^{\beta} \int \overline{ (e^{-ik}-1) \omega_k^{-1} \partial_t \widehat{R}_v(\ell)}
(e^{-ik}-1)\omega_k^{-1} \widehat{R}_v(\ell-\ell_1)\widehat{R}_v(\ell_1) d \ell_1 d \ell \\
&=  2 \varepsilon^{\beta} \int \overline{  \partial_t \widehat{R}_v(\ell)}
 \widehat{R}_v(\ell-\ell_1)\widehat{R}_v(\ell_1) d \ell_1 d \ell \\
&=   \frac23 \varepsilon^{\beta}   \partial_t \int \overline{  \widehat{R}_v(\ell)}
 \widehat{R}_v(\ell-\ell_1)\widehat{R}_v(\ell_1) d \ell_1 d \ell,
\end{align*}  
and similarly $ s_{11} $ as 
$$ 
s_{11} =  \frac23 \varepsilon^{\beta}   \partial_t \int \overline{  \widehat{R}_u(\ell)}
 \widehat{R}_u(\ell-\ell_1)\widehat{R}_u(\ell_1) d \ell_1 d \ell.
$$
Since $s_5 + s_{11} = s_3 + s_9$, the corresponding terms define the other half of $\varepsilon^{\beta} E_2$, where $E_2$ is given by (\ref{E-2-fourier}). \\

 \noindent
vi) The residual terms $ s_6 $ and $ s_{12} $ are estimated with the Cauchy-Schwarz and Young inequalities as
\begin{align*}
|s_6 | &= \varepsilon^{-\beta} \left| \int \overline{\omega_k^{-1}  \partial_t \widehat{R}_u(\ell)} \omega_k^{-1} 
\widehat{\textrm{Res}}_u(\varepsilon^{2} \psi_u, \varepsilon^{2} \psi_v)(\ell) d \ell \right| \\
 &\leq  \varepsilon^{- \beta} \| \omega_k^{-1} \partial_t  \widehat{R}_u \|_{L^2}
 \| \omega_k^{-1} \widehat{\textrm{Res}}_u(\varepsilon^{2} \psi_u, \varepsilon^{2} \psi_v) \|_{L^2} \\
 &\leq  \varepsilon^{3} \| \omega_k^{-1} \partial_t  \widehat{R}_u \|_{L^2}^2+ 
( \varepsilon^{- \beta-\frac{3}{2}} \| \omega_k^{-1}
 \widehat{\textrm{Res}}_u(\varepsilon^{2} \psi_u, \varepsilon^{2} \psi_v)\|_{L^2})^2,
\end{align*}
and similarly, 
\begin{align*}
	|s_{12}| \leq \varepsilon^{3} \| \omega_l^{-1} \partial_t  \widehat{R}_v \|_{L^2}^2  + 
	( \varepsilon^{- \beta-\frac{3}{2}} \| \omega_l^{-1}
 \widehat{\textrm{Res}}_v(\varepsilon^{2} \psi_u, \varepsilon^{2} \psi_v). \|_{L^2})^2
\end{align*}  
By using the estimate (\ref{curry}) and setting $\beta = \frac{5}{2}$, we finally obtain 
\begin{align*}
|s_6|+|s_{12}| \leq \varepsilon^{3} \left( \| \omega_k^{-1} \partial_t  \widehat{R}_u \|_{L^2}^2 + \| \omega_l^{-1} \partial_t  \widehat{R}_v \|_{L^2}^2 + C_{\rm res} \right),
\end{align*}  
for some constant $C_{\rm res} > 0$ that depends on the solution $A$ 
of the KP-II equation (\ref{kp2}).\\

Combining all estimates together, we have derived the energy balance equation 
in the form 
$$ 
\frac{d}{dt} E \leq C_0  \varepsilon^{3} E_0  + C_{\rm res} \varepsilon^{3},
$$ 
for some constant $C_0 > 0$ that also depends on the solution $A$ 
of the KP-II equation (\ref{kp2}), where $E = E_0 + \varepsilon^2 E_1 + \varepsilon^{\beta} E_2$ and $\beta = \frac{5}{2}$. The corrections 
of the leading-order energy $E_0$ are controlled by 
\begin{align*}
|E_1| &\leq 2 (\| \widehat{\psi}_u \|_{L^1} + \| \widehat{\psi}_v \|_{L^1} ) E_0, \\
|E_2| &\leq \frac{2}{3} (\| \widehat{R}_u \|_{L^1} + \| \widehat{R}_v \|_{L^1} ) E_0,
\end{align*}
where 
\begin{align*}
& \| \widehat{\psi}_u \|_{L^1} + \| \widehat{\psi}_v \|_{L^1} \leq 
\sqrt{2\pi} \left( \| \widehat{\psi}_u \|_{L^2} + \| \widehat{\psi}_v \|_{L^2} \right)  \leq C_A, \\
& \| \widehat{R}_u \|_{L^1} + \| \widehat{R}_v \|_{L^1} \leq 
\sqrt{2\pi} \left(  \| \widehat{R}_u \|_{L^2}+ \| \widehat{R}_v \|_{L^2} \right) \leq 2 \sqrt{2\pi E_0},
\end{align*}
with $C_A > 0$ that depends on the solution $A$ of the KP-II equation (\ref{kp2}). As long as there exists $M < \infty$ such that $E_0 \leq M$, there exists $\varepsilon_0 > 0$ such that 
for all 
$ \varepsilon \in (0, \varepsilon_0) $ the energy $E  = E_0 + \varepsilon^2 E_1 + \varepsilon^{\beta} E_2$ is equivalent to the leading-order energy $E_0$, e.g.,
\begin{equation}
\label{equiv}
E_0 \leq 2 E \leq 4 E_0.
\end{equation}
Hence, the energy balance equation can be written as 
$$ 
\frac{d}{dt} E \leq C  \varepsilon^{3} E  + C_{\rm res} \varepsilon^{3},
$$ 
which yields by using Gronwall's inequality for all $ t \in [0,\varepsilon^{-3} \tau_0 ] $ that 
$$ 
E(t) \leq C_{\rm res} \varepsilon^{3} t e^{C \varepsilon^{3}  t} \leq C_{\rm res} \tau_0 e^{C \tau_0} = : \frac{M}{2}.
$$
In view of the equivalence (\ref{equiv}), this verifies that $E_0(t) \leq M$ 
for all $ t \in [0,\varepsilon^{-3} \tau_0 ] $. Due to the scaling (\ref{error-terms}) with $\beta = \frac{5}{2}$, this bound completes the proof of Theorem \ref{th-hor}.

\section{Propagation along an arbitrary direction}
\label{sec3}

Here we consider an arbitrary angle of propagation with respect to 
the square lattice $\Z^2$ and derive the extended KP-II equation 
as the leading-order approximation. This extended KP-II equation is needed to reduce the size of the residual terms and it can be split 
into the sum of the KP-II equation for the main term 
and the linearized KP-II equation for the correction term. 
The approximation theorem is formulated for the smooth solutions 
to the KP-II and linearized KP-II equations. Classes of such smooth solutions are discussed in Section \ref{sec-last}.

\subsection{The formal long-wave limit}

We take the advantage that the Laplacian is invariant under the rotation 
in the plane $\R^2$ and define the leading-order approximation in physical space by 
\begin{eqnarray}
\label{approx2}
	u_{m,n}(t) = \varepsilon^2 A(X,Y,T), \qquad 
	v_{m,n}(t) = \varepsilon^2 B(X,Y,T),
\end{eqnarray}
where
$$
X = \varepsilon( (\cos \phi) m + (\sin \phi) n - t) ,  \quad 
Y = \varepsilon^2 ( - (\sin \phi)m + (\cos \phi)n ) ,  \quad 
T = \varepsilon^3 t , 
$$
and the angle of propagation $\phi \in (0,\frac{\pi}{2})$ determines 
the direction of propagation $ (\cos \phi, \sin \phi)$ with respect to 
the square lattice $\Z^2$. The long wave limit can be written in the extended form compared to system (\ref{start3}) 
and (\ref{start2}):
\begin{equation} 
\label{start5}
\partial_t^2 u = \partial_x^2 u  +  \partial_y^2 u  +  \frac{1}{12} \partial_x^4 u  +  \frac{1}{12} \partial_y^4 u 
-  \partial_x^2 (u^2)  - \partial_x  \partial_y (v^2) 
- \frac{1}{2} (\partial_x - \partial_y) \partial_x \partial_y (v^2) + \mbox{\rm h.o.t.}
\end{equation}
and
\begin{equation}
\label{start4}
\partial_x v + \frac{1}{2} \partial_x^2 v + \mbox{\rm h.o.t.} = \partial_y u + \frac{1}{2} \partial_y^2 u + \mbox{\rm h.o.t.}.
\end{equation}
We find by the chain rule 
\begin{align*}
	\partial_t^2 & = \varepsilon^2 \partial_X^2 - 2 \varepsilon^4 \partial_X \partial_T +  \varepsilon^6 \partial_T^2, \\
	\partial_x^2  & = \varepsilon^2 (\cos \phi)^2  \partial_X^2 - 2 \varepsilon^3  (\cos \phi) (\sin \phi) \partial_X \partial_Y +  \varepsilon^4 (\sin \phi)^2  \partial_Y^2 , \\  
	\partial_y^2 & = \varepsilon^2 (\sin \phi)^2  \partial_X^2 + 2 \varepsilon^3  (\cos \phi) (\sin \phi) \partial_X \partial_Y +  \varepsilon^4 (\cos \phi)^2  \partial_Y^2 , \\
	\partial_x  \partial_y  & =  \varepsilon^2  (\cos \phi)  (\sin \phi) \partial_X^2 +  2 \varepsilon^3 ((\cos \phi)^2-(\sin \phi)^2)\partial_X \partial_Y - \varepsilon^4 (\cos \phi) (\sin \phi ) \partial_Y^2.
\end{align*}
All terms up to the formal order of $\mathcal{O}(\varepsilon^5)$ cancel out, 
whereas the KP-II equation appears at the formal order of $\mathcal{O}(\varepsilon^6)$. For the propagation in the $ x $-direction, 
there are no terms of the formal order of $ \mathcal{O}(\varepsilon^7) $.
If $A$ satisfies the KP-II equation, the residual terms have the formal 
order of $ \mathcal{O}(\varepsilon^8) $. This is no longer the case for the propagation along an arbitrary direction with $\phi\in (0,\frac{\pi}{2})$. 

Substituting (\ref{approx2}) into (\ref{start5}) and (\ref{start4}) 
and removing the terms of the formal order of $ \mathcal{O}(\varepsilon^6) $ and $ \mathcal{O}(\varepsilon^7) $ yield the extended KP-II equation
\begin{align} 
\notag
- 2  \partial_X \partial_T A = & \frac{1}{12} [(\cos \phi)^4 + (\sin \phi)^4] \partial_X^4 A +  \partial_Y^2 A \\ &  - (\cos \phi)^2 \partial_X^2 (A^2) - (\sin \phi) (\cos \phi)) \partial_X^2 (B^2) 
\notag \\
& - \frac{1}{3} \varepsilon [(\cos \phi)^2- (\sin \phi)^2] (\cos \phi)(\sin \phi) \partial_X^3 \partial_Y A  + 2 \varepsilon (\cos \phi) (\sin \phi) \partial_X \partial_Y (A^2) \notag \\
& - \varepsilon [(\cos \phi)^2 - (\sin \phi)^2] \partial_X \partial_Y (B^2) \notag \\
& - \frac{1}{2} \varepsilon [\cos \phi - \sin \phi] (\cos \phi) (\sin \phi) \partial_X^3 (B^2) . \label{kp-approx}
\end{align}
and the relation between the amplitudes $A$ and $B$:
\begin{align*} 
(\cos \phi) \partial_X B - \varepsilon (\sin \phi) \partial_Y B 
+ \frac{1}{2} \varepsilon (\cos \phi)^2 \partial_X^2 B 
= (\sin \phi) \partial_X A  + \varepsilon (\cos \phi) \partial_Y A 
+ \frac{1}{2} \varepsilon (\sin \phi)^2 \partial_X^2 A. 
\end{align*}
Writing 
\begin{align*}
A &= A_1 + \varepsilon A_2, \\
B &= B_1 + \varepsilon B_2,
\end{align*}
we obtain 
\begin{align*}
B_1 &= (\tan \phi) A_1, \\
B_2 &= (\tan \phi) A_2 + \frac{1}{(\cos \phi)^2} \partial_X^{-1} \partial_Y A_1 
+ \frac{1}{2} (\tan \phi) [\sin \phi - \cos \phi] \partial_X A_1, 
\end{align*}
after which the extended KP-II equation (\ref{kp-approx}) can be split into the KP-II equation for $A_1$ 
and the linearized KP-II equation for $A_2$, which are given by 
\begin{align} 
\notag
- 2  \partial_X \partial_T A_1 = & \frac{1}{12} [(\cos \phi)^4 + (\sin \phi)^4] \partial_X^4 A_1 +  \partial_Y^2 A_1 \\ &  - [(\cos \phi)^2 + (\sin \phi)^2 (\tan \phi)] \partial_X^2 (A_1^2) \label{KP}
\end{align}
and
\begin{align} 
- 2  \partial_X \partial_T A_2 = & \frac{1}{12} [(\cos \phi)^4 + (\sin \phi)^4] \partial_X^4 A_2 +  \partial_Y^2 A_2 \notag \\ 
&  - 2 [(\cos \phi)^2 + (\sin \phi)^2 (\tan \phi)] \partial_X^2 (A_1 A_2) \notag \\  &  - \frac{1}{3} [(\cos \phi)^2- (\sin \phi)^2] (\cos \phi)(\sin \phi) \partial_X^3 \partial_Y A_1 \notag \\
& - 2 (\tan \phi)^2 \partial_X^2  (A_1 \partial_X^{-1} \partial_Y A_1) \notag \\
& + [(\sin \phi)^2 (\tan \phi)^2 - (\sin \phi)^2 + 2 (\sin \phi) (\cos \phi) ] \partial_X \partial_Y (A_1^2).  \label{KP2}
\end{align}

\subsection{Estimates for the residual}
\label{compasec}

The leading-order approximation in Fourier space is 
denoted as before by $(\hat{u},\hat{v}) =  (\varepsilon^2 \widehat{\psi_u}, \varepsilon^2 \widehat{\psi_v})$ with 
\begin{equation} 
\label{scal1}
\hat{\psi}_u(k,l,t) = \varepsilon^{-3} e^{i \varkappa t} \chi_{\mathbb{T}^2}(k,l) \widehat{A} ( \varepsilon^{-1} \varkappa, \varepsilon^{-2} \vartheta,T),
\end{equation}
and
\begin{equation} 
\label{scal2}
\hat{\psi}_v(k,l,t) = \varepsilon^{-3} e^{i \varkappa t} \chi_{\mathbb{T}^2}(k,l) \widehat{B} ( \varepsilon^{-1} \varkappa, \varepsilon^{-2} \vartheta,T),
\end{equation}
where 
$$
\varkappa = (\cos \phi) k + (\sin \phi) l, \qquad 
\vartheta = -(\sin \phi) k + (\cos \phi) l,
$$
and 
$$
\hat{A} = \hat{A}_1 + \varepsilon \hat{A}_2, \qquad 
\hat{B} = \hat{B}_1 + \varepsilon \hat{B}_2.
$$

For $\phi \in (0,\frac{\pi}{2})$, the residual terms are formally given in physical space by 
\begin{align*}
{\textrm{Res}}_u(\varepsilon^2 \psi_u,\varepsilon^2 \psi_v) & = 
	\mathcal{O}( \varepsilon^8 \partial_T^2 A_1 , 
	\varepsilon^8  \partial_X^6 A_1,
	\varepsilon^8  \partial_X^2 \partial_Y^2 A_1 , 
\varepsilon^8   \partial_X^4 (A_1^2) ,   \varepsilon^8   \partial_Y^2 (A_1^2) , \varepsilon^8   \partial_Y^2 (B_1^2) , \\
& \qquad \varepsilon^9 \partial_T^2 A_2 , 
\varepsilon^9  \partial_X^6 A_2,
\varepsilon^9  \partial_X^2 \partial_Y^2 A_2 , 
	\varepsilon^8   \partial_X^2 (A_2^2) ,  \varepsilon^8   \partial_X \partial_Y (A_1 A_2) , 	\\
& \qquad \varepsilon^8   \partial_X^2 (B_2^2),  \varepsilon^8   \partial_X \partial_Y (B_1 B_2) , \varepsilon^8 \partial_X^3 (B_1^2) )
\end{align*}
and 
\begin{align*}
{\textrm{Res}}_v(\varepsilon^2 \psi_u,\varepsilon^2 \psi_v) & = 
\mathcal{O}( \varepsilon^8 \partial_T^2 B_1 , 
\varepsilon^8  \partial_X^6 B_1,
\varepsilon^8  \partial_X^2 \partial_Y^2 B_1 , 
\varepsilon^8   \partial_X^4 (B_1^2) ,   \varepsilon^8   \partial_Y^2 (B_1^2) , \varepsilon^8   \partial_Y^2 (A_1^2) , \\
& \qquad \varepsilon^9 \partial_T^2 B_2 , 
\varepsilon^9  \partial_X^6 B_2,
\varepsilon^9  \partial_X^2 \partial_Y^2 B_2 , 
\varepsilon^8   \partial_X^2 (B_2^2) ,  \varepsilon^8   \partial_X \partial_Y (B_1 B_2) , 	\\
& \qquad \varepsilon^8   \partial_X^2 (A_2^2),  \varepsilon^8   \partial_X \partial_Y (A_1 A_2) , \varepsilon^8 \partial_X^3 (A_1^2) ).
\end{align*}

The residual terms containing $A_1$ are controlled by the local well-posedness theory of Lemma \ref{KPIIwellpos2} with $A_1$ being a smooth solution of the KP-II equation (\ref{KP}). If $A_2$ enjoys the same properties, 
the bound (\ref{curry}) is justified, from which the proof of the approximation theorem stated below is analogous to the proof of Theorem \ref{th-hor}.

\begin{theorem}
	\label{th-arb}
	There exist  $ C_0 $ and $ \varepsilon_0 > 0 $ such that for all $ \varepsilon  \in (0, \varepsilon_0) $ the following holds. Let $ A_1 \in 
	C([0,\tau_0], H^{s+9})  $ be a solution of the KP-II equation \eqref{KP} given by Lemma \ref{KPIIwellpos2} with fixed $s \geq 0$ and assume 
	that $ A_2 \in C([0,\tau_0], H^{s+9})  $ is a solution of the linearized KP-II equation \eqref{KP2} with the same properties as for $A_1$. Then there exist solutions $ (u,v) $ of system \eqref{umneq} with 
	$$
	\sup_{t \in [0,\varepsilon^{-3} \tau_0]} \| u(t) - \varepsilon^2 \psi_u(t) \|_{\ell^2(\Z^2)} + \| v(t) - \varepsilon^2 \psi_v(t) \|_{\ell^2(\Z^2)}  \leq C_0 \varepsilon^{\frac{5}{2}},
	$$
	where $(\psi_u,\psi_v)$ are given by the inverse Fourier transform 
	of (\ref{scal1}) and (\ref{scal2}).
\end{theorem}

\section{Discussion}
\label{sec-last}

Here we discuss classes of solutions to the KP-II equation (\ref{KP}) 
and the linearized KP-II equation (\ref{KP2}) for which the approximation 
result of Theorem \ref{th-arb} can be obtained. This includes transversely independent solutions, periodic solutions, and decaying solutions in the unbounded domain. 

\subsection{Transversely independent solutions}

Let $A_{1,2} = A_{1,2}(X,T)$ be $Y$-independent. Then, $A_1$ is a solution 
of the KdV equation 
\begin{align} 
- 2  \partial_T A_1 = \frac{1}{12} [(\cos \phi)^4 + (\sin \phi)^4] \partial_X^3 A_1 - [(\cos \phi)^2 + (\sin \phi)^2 (\tan \phi)] \partial_X (A_1^2) \label{KdV}
\end{align}
and $A_2$ is a solution of the linearized KdV equation
\begin{align} 
- 2 \partial_T A_2 = \frac{1}{12} [(\cos \phi)^4 + (\sin \phi)^4] \partial_X^3 A_2 - 2 [(\cos \phi)^2 + (\sin \phi)^2 (\tan \phi)] \partial_X (A_1 A_2).  \label{linKdV}
\end{align}
Clearly, if $A_2 |_{T = 0} = 0$, then $A_2(X,T) \equiv 0$. 
Smooth solutions for $A_1$ to the KdV equation (\ref{KdV}) exist 
in one-dimensional Sobolev spaces without any additional constraints on the initial data $A_1 |_{T = 0} \in H^s(\R)$. As a result, the approximation result of Theorem \ref{th-arb} holds in one-dimensional Sobolev spaces $H^s(\R)$ for every $s \geq 6$. 

\subsection{Periodic solutions}

Let $A_{1,2}$ be spatially periodic in $X$ and $Y$. Without loss of generality, 
we assume that the solutions are $1$-periodic in $X$ and $Y$. 
Such solutions can be expressed in the Fourier form, e.g.
$$ 
A(X,Y,T)  = \sum_{j_1 \in \Z} \sum_{j_2\in \Z} \widehat{A}_{j_1,j_2}(T) e^{2 i \pi ( j_1 X +  j_2 Y)},
$$
with $ \widehat{A}_{j_1,j_2}(T) \in \C $. The Fourier transformed KP equation (\ref{KP}) vanishes identically for $ j_1 = j_2 = 0 $ and so there is no governing equation for $ \widehat{A}_{0,0}(T) $. Therefore, we are free to set $ \widehat{A}_{0,0}(T) \equiv 0$. Global well-posedness of the KP-II equation \eqref{KP} was established in $H^s( \mathbb{T}^2)$ for any $s\geq 0$ in \cite{Bourgain}, provided that the initial data satisfies 
\begin{equation}
\label{mean-zero}
\oint A(X,Y,0) dX = 0, \quad \mbox{\rm for every  } Y,
\end{equation}
that is, $\widehat{A}_{0,j_2} = 0$ for every $j_2 \in \Z$.

The evolution problem for the inhomogeneous linearized KP equation (\ref{KP2})
is also well-defined in Sobolev spaces $H^s(\mathbb{T}^2)$ with the same constraint (\ref{mean-zero}). Thus, antiderivatives in $X$ presents no problem 
on existence of smooth solutions in Sobolev spaces $H^s(\mathbb{T}^2)$ both 
for the KP-II equation (\ref{KP}) and the linearized KP-II equation (\ref{KP2}).

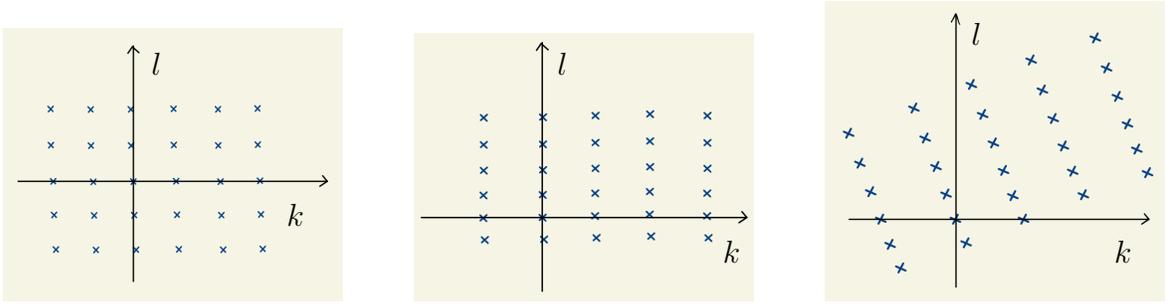
\begin{figure}[htbp] %  figure placement: here, top, bottom, or page
	\centering

	\includegraphics[width=4.5cm]{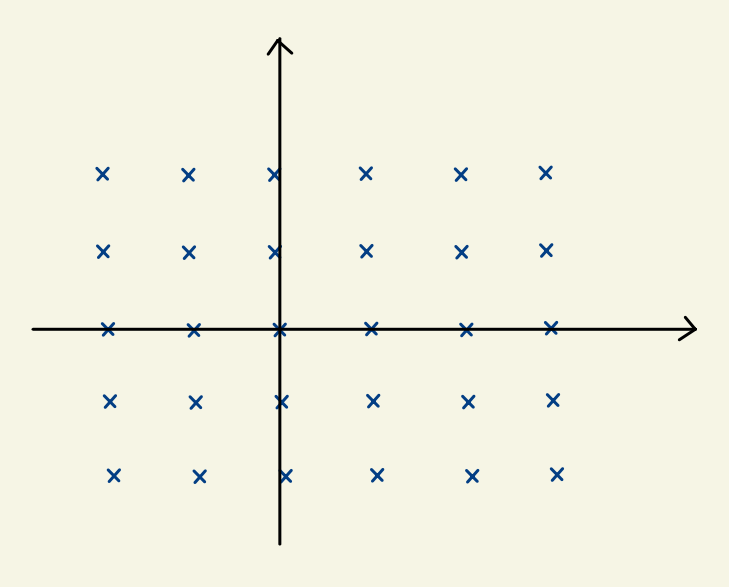} \qquad
	\includegraphics[width=4.5cm]{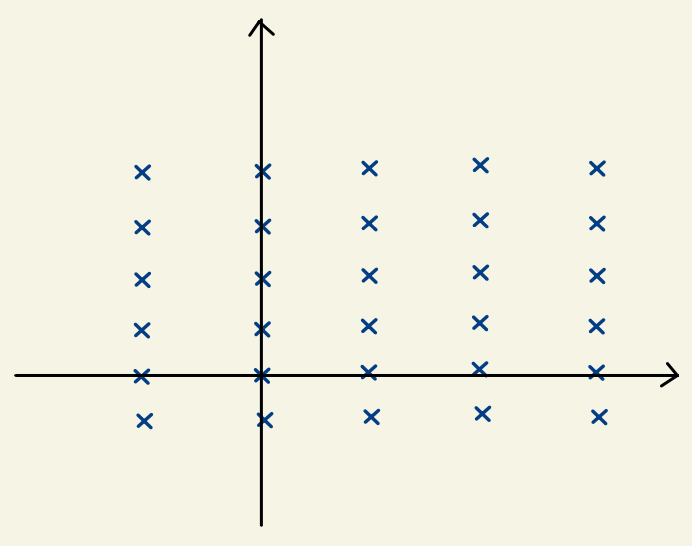} \qquad
	\includegraphics[width=4.5cm]{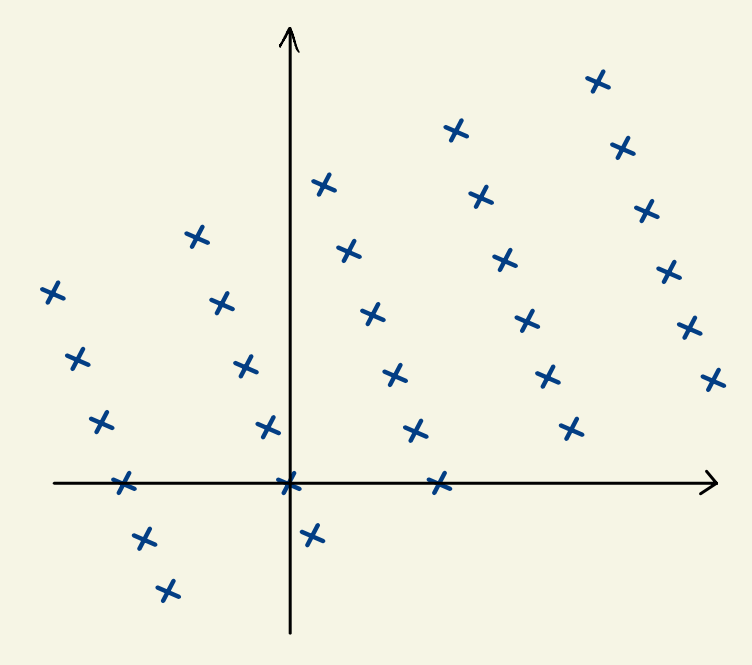}
	\begin{picture}(14,6)
	\setlength{\unitlength}{1cm}
	\put(-0.8, 0.5){$ k $}
	\put(-2.7, 3.4){$ l $}
	
	\put(-6, 0.5){$ k $}
	\put(-8.2, 3){$ l $}

	\put(-11.8, 1){$ k $}
	\put(-13.6, 3){$ l $}
	\end{picture}
	
	\caption{The left panel shows the distribution of Fourier modes in the case of periodic boundary conditions for the KP equation (\ref{KP}). 
		The dots are located at integer multiples of $ 2 \pi $. 
		The middle panel shows the resulting distribution of Fourier modes in the original system (\ref{umneq}) in the case of wave propagation along the $ x $-axis. The distance of the dots in $ k $-direction is $ \mathcal{O}(\varepsilon) $ and in 
		$ l $-direction $  \mathcal{O}(\varepsilon^{2}) $.
		The right panel shows the corresponding distribution of Fourier modes in the original system (\ref{umneq}) in case of wave propagation along an arbitrary direction. For a rational propagation direction $(\cos \phi,\sin \phi)$, we obtain a periodic lattice in Fourier space with finitely many modes on the torus. For an irrational propagation direction $(\cos \phi,\sin \phi)$, we obtain a quasi-periodic lattice that lies densely in the torus due to non-linear interactions.}
	\label{figmodesperiodic}
\end{figure}

However, the choice of periodic boundary conditions in the KP equation (\ref{KP}) leads to the problem of having to choose corresponding boundary conditions in the original FPU system (\ref{umneq}).
The resulting difficulties are illustrated in Figure \ref{figmodesperiodic}.
An irrational propagation direction $(\cos \phi,\sin \phi)$ leads to a quasi-periodic lattice in Fourier space which lies densely in the torus. 
The treatment of this problem leads to functional analytical difficulties whose solution is outside the scopes of this work.
Therefore, in the following we restrict ourselves to the case of rational propagation directions $(\cos \phi,\sin \phi)$.

In case of a propagation along the $ x $-axis 
spatially periodic solutions for the KP equation (\ref{kp2}) of period $ 1$ in $ X $ and $ Y $ correspond in the original FPU system \eqref{umneq} to 
$$
u_{m,n}  = u_{m+1 /\varepsilon,n} = u_{m,n+1 /\varepsilon^2},
$$ 
and so we should restrict to values of $ \varepsilon  > 0 $ such that 
$ 1/\varepsilon \in \N $ for which $ 1/\varepsilon^2 \in \N $.
Such solutions can be represented by the finite Fourier polynomial
$$ 
u_{m,n}  = \sum_{j_1= -1/\varepsilon}^{1/\varepsilon-1} \sum_{j_2= -1/\varepsilon^2}^{1/\varepsilon^2-1} \widehat{u}_{j_1,j_2} e^{2 \pi i(j_1 \varepsilon m)} e^{2 \pi i(j_2 \varepsilon^2 n)}
$$ 
In case of a rational direction of propagation $ (\cos \phi, \sin \phi)$ with respect to 
the square lattice $\Z^2$ the  finite Fourier polynomial is given by
\begin{eqnarray*}
	u_{m,n} &  = &\sum_{j_1= -1/\varepsilon}^{1/\varepsilon-1} \sum_{j_2= -1/\varepsilon^2}^{1/\varepsilon^2-1} \widehat{u}_{j_1,j_2} e^{2 \pi i(j_1 \varepsilon ( (\cos \phi) m + (\sin \phi) n ) )} e^{2 \pi i(j_2 \varepsilon^2 ( - (\sin \phi)m + (\cos \phi)n ))} \\ 
	&  = &
	\sum_{j_1= -1/\varepsilon}^{1/\varepsilon-1} \sum_{j_1= -1/\varepsilon^2}^{1/\varepsilon^2-1} \widehat{u}_{j_1,j_2} e^{2 \pi i ( j_1 \varepsilon (\cos \phi) -j_2 \varepsilon^2   (\sin \phi))m  } e^{2 \pi i (j_1 \varepsilon(\sin \phi)  +  j_2 \varepsilon^2  (\cos \phi))n } .
\end{eqnarray*}
Since the indices do not reflect the position in Fourier space we introduce 
$$ 
\widehat{u}(2 \pi  ( j_1 \varepsilon (\cos \phi) -j_2 \varepsilon^2   (\sin \phi)),2 \pi  (j_1 \varepsilon(\sin \phi)  +  j_2 \varepsilon^2  (\cos \phi))) =  \widehat{u}_{j_1,j_2} 
$$ 
and introduce the lattice
$$ 
\Sigma_2 = \{ (2 \pi  ( j_1 \varepsilon (\cos \phi) -j_2 \varepsilon^2   (\sin \phi)),2 \pi  (j_1 \varepsilon(\sin \phi)  +  j_2 \varepsilon^2  (\cos \phi))) \in [0,2\pi)^2 : j_1,j_2 \in \Z \}
$$ 

The essential difference to the above calculations is that the space $ L^2(\mathbb{T}^2) $ has to be replaced by a sequence space, i.e., 
instead of $ L^2(\mathbb{T}^2) $  for solving the FPU system in Fourier space 
we consider the space $ \ell^2(\Sigma_2) $ which is equipped with the norm
$$ 
\| \widehat{u} \|_{\ell^2(\Sigma_2)} = \varepsilon^{3/2} \left(
\sum_{j_1= -1/\varepsilon}^{1/\varepsilon-1} \sum_{j_2= -1/\varepsilon^2}^{1/\varepsilon^2-1} |\widehat{u}_{j_1,j_2}|^2\right)^{1/2} = \varepsilon^{3/2} \left(\sum_{\ell \in \Sigma_2} |\widehat{u}(\ell)|^2\right)^{1/2}
$$ 
in order to have the same scaling  as above.
For estimating the leading-order approximation 
$ (\hat{u},\hat{v}) = (\varepsilon^2 \widehat{\psi}_u, \varepsilon^2 \widehat{\psi}_v) $ in
the equations for the error
we need similar to  above a
space $ \ell^1(\Sigma_2) $ which is equipped with the norm
$ 
\| \widehat{u} \|_{\ell^1(\Sigma_2)} = \varepsilon^{3} \sum_{\ell \in \Sigma_2} |\widehat{u}(\ell)| 
$. 
By Cauchy--Schwarz inequality, we obtain 
$$
\| \widehat{u} \|_{\ell^1(\Sigma_2)} = \varepsilon^{3} \sum_{\ell \in \Sigma_2} 1 \cdot |\widehat{u}(\ell)| \leq \varepsilon^{3} (\sum_{\ell \in \Sigma_2} 1^2  )^{1/2}
(\sum_{\ell \in \Sigma_2} |\widehat{u}(\ell)|^2)^{1/2} \leq  C \| \widehat{u} \|_{\ell^2(\Sigma_2)}
$$
where we used  $ \sum_{\ell \in \Sigma_2} 1^2 = \mathcal{O}(\varepsilon^{-3}) $ due to the $ \mathcal{O}(\varepsilon^{-3}) $ many summands.
As a result, we have
$ \| \widehat{\psi}_u \|_{\ell^1(\Sigma_2)} = \mathcal{O}(1)  $ and
$ \| \widehat{\psi}_v \|_{\ell^1(\Sigma_2)} = \mathcal{O}(1) $. 
With these norms and notations the approximation result of Theorem \ref{th-arb} transfers to the periodic Sobolev spaces $H^s(\mathbb{T}^2)$ for every $s \geq 9$.

\subsection{Unbounded domain}

One needs to be careful in analyzing smooth solutions 
of the KP-II equation (\ref{KP}) and the linearized KP-II equation (\ref{KP2}) 
in Sobolev spaces $H^s(\mathbb{R}^2)$. As was pointed out in \cite{Molinet}, if $A$ belongs to $H^s(\R^2)$, then $\partial_X^{-1} \partial_Y A^2$ may not be in $H^s(\R^2)$ as the integral of the positive function 
cannot decay to zero both as $X \to -\infty$ and $X \to +\infty$.  
This was also observed in the proof of Lemma 1 in \cite{niki}, where a constraint was added on the combined quantity $\partial_X^{-1}\partial_Y^2 ( \partial_X^{-2}\partial_Y^2 A_0 - A_0^2)$ rather than on $\partial_X^{-3} \partial_Y^3 A$ or $\partial_X^{-1} \partial_Y A^2$.

Rewriting the linearized KP-II equation (\ref{KP2}) in the evolution form 
yields 
\begin{align} 
- 2  \partial_T A_2 = & \frac{1}{12} [(\cos \phi)^4 + (\sin \phi)^4] \partial_X^3 A_2 +  \partial_X^{-1} \partial_Y^2 A_2 \notag \\ 
&  - 2 [(\cos \phi)^2 + (\sin \phi)^2 (\tan \phi)] \partial_X (A_1 A_2) \notag \\  &  - \frac{1}{3} [(\cos \phi)^2- (\sin \phi)^2] (\cos \phi)(\sin \phi) \partial_X^2 \partial_Y A_1 \notag \\
& - 2 (\tan \phi)^2 \partial_X  (A_1 \partial_X^{-1} \partial_Y A_1) \notag \\
& + [(\sin \phi)^2 (\tan \phi)^2 - (\sin \phi)^2 + 2 (\sin \phi) (\cos \phi) ] \partial_Y (A_1^2).  \label{linKP}
\end{align}
In the evolution form, the right-hand side of the linearized KP-II equation 
contains terms $\mathcal{O}(\partial_X^2 \partial_Y A_1, \partial_X(A_1 \partial_X^{-1} \partial_YA_1),\partial_Y (A^2_1))$, which are controlled by 
Lemma \ref{KPIIwellpos2} in Sobolev norms. Therefore, by Duhamel's principle, we have $A_2 \in C([0,\tau_0],H^{s+6}) \cap C^1([0,\tau_0],H^{s+3})$. However, for the justification analysis, 
we need to estimate $\partial_X^{-1} \partial_T^2 A_2$ in Sobolev norms. 

For $D_2 := \partial_X^{-1} \partial_YA_2$, we can obtain from (\ref{linKP})
\begin{align*} 
- 2  \partial_T D_2 = & \frac{1}{12} [(\cos \phi)^4 + (\sin \phi)^4] \partial_X^3 D_2 +  \partial_X^{-1} \partial_Y^2 D_2\\ 
&  - 2 [(\cos \phi)^2 + (\sin \phi)^2 (\tan \phi)] \partial_Y (A_1 A_2)  \\  &  - \frac{1}{3} [(\cos \phi)^2- (\sin \phi)^2] (\cos \phi)(\sin \phi) \partial_X \partial_Y^2 A_1 \\
& - 2 (\tan \phi)^2 \partial_Y  (A_1 \partial_X^{-1} \partial_Y A_1)  \\
& + [(\sin \phi)^2 (\tan \phi)^2 - (\sin \phi)^2 + 2 (\sin \phi) (\cos \phi) ] \partial_X^{-1} \partial_Y^2 (A_1^2).
\end{align*}
By Duhamel's principle, we obtain 
$\partial_X^{-1} \partial_Y A_2 \in C^0([0,\tau_0],H^{s+6}) \cap C^1([0,\tau_0],H^{s+3})$ if the initial data satisfy the constraint 
$$
\partial_X^{-1} \partial_Y^2 \left[ \partial_X^{-1} \partial_Y A_2 |_{T=0} 
+ [(\sin \phi)^2 (\tan \phi)^2 - (\sin \phi)^2 + 2 (\sin \phi) (\cos \phi) ] A_1^2 |_{T=0} \right] \in H^{s+6 }.
$$
Since we need to control $\partial_X^{-1} \partial_T^2 A_2$, we need 
to extend this method to $\partial_X^{-1} \partial_Y A_2 \in C^2([0,\tau_0],H^s)$ or equivalently to 
$\partial_X^{-2} \partial_Y^2 A_2 \in C^0([0,\tau_0],H^{s+3}) \cap C^1([0,\tau_0],H^{s})$. However, this is out of reach at the present time 
because of overdetermined set of constraints on the initial data 
$A_1 |_{T=0}$ and $A_2 |_{T=0}$. Further work on extending the well-posedness results for the KP-II equation (\ref{KP2}) is needed to satisfy 
the requirements of Theorem \ref{th-arb} in $H^s(\R^2)$ for every $s \geq 9$.

\bibliographystyle{alpha}
%\bibliographystyle{plain} % zum durchz�hlen
%\bibliography{KPbib}

\end{document}